\documentclass{ws-procs9x6}

\usepackage{graphicx}
\usepackage{epsfig,latexsym}
\newcommand{\be}{\begin{eqnarray}}
\newcommand{\ee}{\end{eqnarray}}

\begin{document}

\title{Heavy Quarkonia Above Deconfinement}

\author{\'A. M\'ocsy}

\address{RIKEN-BNL Research Center, Brookhaven National Laboratory\\
Upton, NY 11973, USA\\
E-mail: mocsy@bnl.gov}

\begin{abstract}
In this talk I summarize our current understanding of quarkonium states above deconfinement based on phenomenological and lattice QCD studies. 
\end{abstract}

\keywords{Quarkonium. Deconfinement. Finite temperature QCD. Potential models.}

\section*{}

{\bf Quarkonium as a Signal of Deconfinement.} Quarkonium, the common name of a meson state made by a heavy quark $Q$ and antiquark $\bar{Q}$ has been in the center of interest since the discovery of the $J/\psi$ at BNL and SLAC in 1974. These heavy mesons allowed for a careful testing of Quantum Chromodynamics (QCD), and are essential diagnostic tools of the deconfinement transition in hot QCD. 

The 20 year old prediction by Matsui and Satz\cite{Matsui:1986dk}, that the melting of heavy quark-antiquark bound states at the deconfinement temperature could be considered an unambiguous signal for deconfinement, has led to an intense line of studies.  It was predicted\cite{Matsui:1986dk} that color screening in a deconfined medium causes the dissolution of the $J/\psi$. The main idea behind this prediction is that in the deconfined phase of QCD, refered to as quark-gluon plasma, there is a screening of the color force between a heavy quark and antiquark, much like the known Debye-screening in QED. This screening is due to the light quarks, antiquarks and gluons present in the plasma. The range of screening is characterized by the screening length, which is dependent on temperature. When the screening length is smaller than the size of the boundstate then this will dissociate. In case of the $J/\psi$ this would happen at $T_c$, the temperature of deconfinement. The melting of this state would in turn manifests in the suppression of the $J/\psi$ peak in the dilepton spectrum, providing an experimentally detectable signal for deconfinement. Experiments colliding heavy ions at ultrarelativistic energies at SPS-CERN and RHIC-BNL have been looking for $J/\psi$ suppression\cite{experiment}.
 
It is clear that understanding the modification of the properties of the $J/\psi$, and other quarkonium states in a hot medium is therefore crucial for understanding deconfinement. Until recently, theoretical studies, including the one the original prediction is based on, have been mostly phenomenological, using potential models as the basic tool\cite{Matsui:1986dk,{Karsch:1987pv},{Digal:2001ue}}. Today, first principle numerical calculations of QCD carried out on the lattice are also available, and provide new information about quarkonia at high temperatures\cite{Umeda:2002vr,{Datta:2003ww},{Petrov:2005ej}}. In what follows, I first discuss potential models and lattice results, then point out the agreements and inconsistencies between these. 

{\bf Potential Models. }
The essence of potential models is to assume that the interaction between a heavy quark and antiquark is mediated by a potential. This assumption is feasible when the $Q-\bar{Q}$ interaction is instantaneous. This adiabatic approximation is applicable in the nonrelativistic situation, when the timescale associated with the relative motion of the $Q$ and $\bar{Q}$ is much larger than that associated with the gluons. The bound state properties are determined by solving the Schr\"odinger equation with the potential. At zero temperature the phenomenologically introduced Cornell potential\cite{Eichten:1974af}, made of a Coulomb and a linear part, was found to describe quarkonia spectroscopy rather well. It was understood much later, that the existence of the hierarchy of energy scales, $m\gg mv\gg mv^2$, in nonrelativistic systems of mass $m$ and velocity $v$,  allows for the construction of a sequence of effective field theories: nonrelativistic QCD (NRQCD) and potential NRQCD (pNRQCD)\cite{Bali:2000gf}.  The Cornell potential can be obtained as the leading order approximation of pNRQCD. The matching between QCD and these effective theories has been done on a perturbative level, but by emphasizing the importance of non-perturbative effects the validity of the potential model approach has been early on challenged by\cite{Voloshin:1978hc}.  
 
Potential models have been used also at finite temperature in context of deconfinement. At finite temperature the form of the potential is not known. A sequence of effective field theories that might yield a potential have not yet been derived from hot QCD. Such a calculation is intrinsically more involved than that at zero temperature, due to the presence of additional scales  determined by the temperature, $T, gT, g^2T$.  So at finite $T$ it is more difficult to establish a well-defined potential. It is thus questionable whether a temperature-dependent potential is adequate for understanding the properties of quarkonia in a hot medium. 

Nevertheless, in lack of other theoretical means potential models have been widely used to investigate quarkonium states at finite temperature. A usual choice for the potential is a screened version of the Cornell potential with a temperature-dependent screening. The screening radius can be determined exactly in perturbative analysis, as the inverse of the Debye-mass $\sim gT$. Given the small size of the quarkonium states one can argue that at the relevant short distances perturbative calculations are acceptable. In finite temperature lattice QCD the potential between a static heavy quark and antiquark has been identified from Polyakov loop correlations measuring the free energy, from which the color screening has also been obtained. Solving the Schr\"odinger equation provides the temperature-dependence of the quarkonium properties, allowing to monitor at what temperature the screening radius drops below the radius of the bound state. The prediction is that the $J/\psi$ disappears around $1.1T_c$\cite{Digal:2001ue}. Since the excited states have a larger radius than the ground state, a pattern of sequential suppression can be identified, meaning that higher excited states dissolve earlier. This sequential effect is predicted to be seen in the $J/\psi$ suppression pattern as a function of the energy density\cite{Digal:2001ue} as determined in experiments\cite{experiment}.

{\bf Quarkonium from Lattice.}
A few years ago an alternative way to study the temperature dependence of quarkonia properties became available in the form of numerical simulations of QCD carried out on a lattice. Correlation functions of hadronic currents are reliably calculated in Euclidean time $\tau$. The spectral representation of the correlator $G(\tau,T)=\int d\omega \sigma(\omega,T) K(\tau,\omega,T)$ allows for the reconstruction of the meson spectral function $\sigma(\omega,T)$ using the Maximum Entropy Method. 
Correlators can be conveniently used to determine the modification of the quarkonium properties in a hot medium at temperatures above $T_c$. To eliminate the trivial T-dependence of the kernel $K(\tau,\omega,T)$ the ratio of the correlator to the so-called reconstructed one, $G_{recon}(\tau,T)=\int d\omega \sigma(\omega,T=0)K(\tau,\omega,T)$ is determined. Any deviation from one of the ratio $G/G_{recon}$ indicates modification of the spectral function with temperature. 

The first results\cite{Umeda:2002vr,{Datta:2003ww}} were obtained in quenched simulations, i.e.~a purely gluonic background, in which the $Q-\bar{Q}$ is at rest with respect to the thermal medium. To the surprise of the community, the data contradicts what has been theoretically expected from the potential model calculations: The ratio of correlators for the ground state charmonium are technically flat even at temperatures well above $T_c$, indicating that the 1S $J/\psi$ and $\eta_c$ survive at least up to 1.5 $T_c$.  The correlator of the 1P state increases for all $T>T_c$, suggesting that the $\chi_c$ dissolves by 1.16 $T_c$. The corresponding spectral functions not only reinforce these findings, but also indicate that the properties of the 1S states, the mass and within errors the amplitude do not change up to these temperatures\cite{Datta:2003ww}.
Although the temperature-dependence of the ground state charmonium correlators was found to be small, a small difference between the behavior of the $J/\psi$ and $\eta_c$ correlators has been identified. Such a difference was a priori not expected. Since hyperfine splitting is not taken into account explicitly, the source for this must be elsewhere. The findings for bottomonium are similar\cite{Petrov:2005ej}: no modification of the ground state, $\Upsilon$ and $\eta_b$, up to temperatures of $2.3T_c$, while dissolution of the $\chi_b$ already at $1.15T_c$. Such early melting of the $\chi_b$ is in contrast with expectations based on the screening argument. The size of this state is similar to that of the $J/\psi$, so similar dissolution temperatures were expected. Calculations performed in two-flavor QCD\cite{Aarts:2005fx} sofar support the qualitative findings of the quenched results. Examining the behavior of quarkonium in motion with respect to the heat bath have also began. The first results\cite{Datta:2004js} show modifications of the finite momenta charmonium when inserted in a deconfined gluonic medium. 

{\bf Potential Models Revisited.}
After the appearance of the lattice data on quarkonium, potential models have been reconsidered, using different temperature dependent potentials\cite{Shuryak:2003ty,{Wong:2004zr},{Alberico:2005xw}}. The internal energy of a $Q-\bar{Q}$ pair as determined on the lattice\cite{Kaczmarek:2003dp} and identified as the potential\cite{Shuryak:2003ty} is used to study the possibility of strongly coupled Coulomb bound states. One should be aware though that in leading order perturbation theory, which is valid at high temperatures, the potential is equal to the free energy of the quark-antiquark pair. Beyond leading order there is an entropy contribution to the free energy which determines the internal energy. There is a sharp peak in the entropy near $T_c$\cite{Kaczmarek:2005uv} making the identification of the internal energy as potential conceptually difficult\cite{Petreczky:2005bd}.  It is exactly this entropy contribution however that makes the internal energy a deeper potential than the free enregy, allowing thus for some of the quarkonium states to remain bound up to temperatures beyond $T_c$. Because of this, the lattice internal energy remains a popular choice for the potential\cite{Alberico:2005xw}. A combination of the lattice internal and free energy has also been suggested as potential\cite{Wong:2004zr}.  A common feature of these potentials is that they incorporate temperature-dependent screening and yield quarkonium dissociation temperatures in accordance with the above quoted numbers from the lattice. 

{\bf Correlators and Spectral Functions.}
We performed the first phenomenological study of quarkonium correlators\cite{Mocsy:2004bv,{Mocsy:2006zd}} using potential models and pointed out that although the agreement of dissociation temperatures from potential models with lattice data is necessary, it is not sufficient to claim understanding the disappearance of quarkonia in the quark-gluon plasma: Even though potential models with certain screened potentials can reproduce qualitative features of the lattice spectral function, such as the survival of the ground state and the melting of the excited states, the temperature dependence of the quarkonium correlators, especially in the S-channel is not reproduced. Furthermore, the properties of the states determined with these screened potentials do not seem to reproduce the results indicated by the lattice spectral functions. 

As mentioned above, solving the Schr\"odinger equation yields the temperature-dependence of the quarkonium  mass, size, and amplitude. It does not however give the spectral function and the correlator. This needs to be determined by  different means. We considered two appraches. One is to design the spectral function as the sum of bound state/resonance contributions and the perturbative continuum above a threshold. This threshold is determined by the asymptotic value of the potential, and decreases with increasing temperature. To make direct comparison with the lattice data we calculate the ratio of correlators $G/G_{recon}$. The behavior for the $\chi_c$ correlator qualitatively agrees with what is seen on the lattice. But there is no agreement for the $\eta_c$ correlator\cite{Mocsy:2004bv} . In the model calculations one can identify a more complex substructure in the $\eta_c$ correlator: The reduction of the continuum threshold and that the amplitude of the states are distinguishable contributions\cite{Mocsy:2004bv} . The small difference between the $J/\psi$ and $\eta_c$ correlators detected on the lattice can also be seen in the model calculations. The difference is attributed to the transport contribution additional in the vector channel compared to the pseudoscalar channel\cite{Petreczky:2005nh}. The increase of the $\chi_b$ correlator is in accordance with lattice results, and is explained as the effect of the decreasing continuum threshold. All these conclusions are unchanged for the different potentials. 

In another approach we perform a full non-relativistic calculation of the Green's function\cite{{Mocsy:2006zd},new}, whose imaginary part provides the quarkonium spectral function. The main advantage of this approach is that resonances and continuum are considered together. The behavior of the ratio of S-wave correlators obtained for the different lattice fitted potentials again are not flat, indicating that the spectral functions of the $\eta_c$, $J/\psi$ and $\Upsilon$ are significantly different than at zero temperature. The spectral function of these states are severely modified in this approach\cite{new}, which is not supported by the lattice results. \\

{\bf Possible Improvements.}
Since none of the popular lattice fitted potentials lead to results in agreement with the lattice, it is  reasonable to repeat the question whether such temperature-dependent screened potentials are the right way to describe modification of quarkonia properties with temperature?  If not, then what is the mechanism responsible for the dissociation of quarkonia at high temperatures? If yes, then what is the potential that can reproduce the data for S and P charmonium and bottomonium states simultaneously?

As a first attempt to answer this question we consider a simple toy model\cite{Mocsy:2006wg}: Because no modification in the properties of the ground state charmonium compared to the zero temperature values has been observed up to well above $T_c$,  keep these at their Particle Data Group values. Plus, since higher excited states seem to disappear near $T_c$, we "melt" these by removing them from the design spectral function at $T_c$.  The main idea is to compensate for the melting of the higher excited states above $T_c$ with the decrease of the only parameter, the continuum threshold. With such a simple model, that does not include temperature dependent screening we were able to recover the flatness of the $\eta_c$ and the increase of the $\chi_c$ correlator. 

Another attempt is to use a "screened" Cornell potential as input to determine the nonrelativistic Green's function. Here "screening" is a parameter that is related neither to Debye-screening, nor has anything to do with the free- and internal energies determined on the lattice. Our preliminary finding is that we can tune this parameter such that some qualitative agreement for the S-waves can be obtained. Determining  the P-wave correlator is part of our currently ongoing research.      

{\bf Final Remaks.} 
We do not yet posses a comprehensive phenomenological tool that can explain consistently all the lattice observations on heavy quarkonium. It is not clear at the moment to what extent the modification of the quarkonium spectral functions can be understood in terms of the color Debye-screening picture. When the time-scale of screening is not short compared to the time-scale of the heavy quark motion, the role of gluo-dissociation in the medium modification of the spectral functions has to be understood. This is a topic of our ongoing investigations. 

{\bf Acknowledgments.}
This presentation is based on work done in collaboration with P.~Petreczky. I thank the Organizers for a successful workshop and J.~Casalderrey-Solana, D.~Kharzeev and H.~Satz for helpful discussions. 

%%%%%%%%%%%%% 

\end{document}